\documentclass[preprint,11pt]{elsarticle}

\usepackage[
  left=3cm,
  right=3cm,
  top=4cm,
  bottom=4cm
]{geometry}

\usepackage{amssymb}
\usepackage{amsmath}

\usepackage{orcidlink}

\journal{Journal of Subatomic Particles and Cosmology}

\usepackage{subcaption}

\newcommand{\ttbar}{t\bar{t}}
\newcommand{\nchel}{Nc_{\rm hel}}

\begin{document}

\begin{frontmatter}



\title{Reconstructing Toponium using Recursive Jigsaw Reconstruction} 

\author[]{Aman Desai \orcidlink{0000-0003-2631-9696}}\ead{aman.desai@adelaide.edu.au}
\author[]{Amelia Lovison \orcidlink{0009-0009-5196-4647}}\ead{amelia.lovison@adelaide.edu.au}
\author{Paul Jackson \orcidlink{0000-0002-0847-402X}}
\ead{p.jackson@adelaide.edu.au}

\affiliation{Department of Physics, The University of Adelaide, North Terrace, Adelaide, SA 5005, Australia}

\begin{abstract}

The results from the ATLAS and CMS experiment at the Large Hadron Collider indicate the existence of a top-quark pair bound state near the $\ttbar$ threshold region. We present a method relying on Recursive Jigsaw Reconstruction to reconstruct the toponium bound state at the $\ttbar$ threshold region. We propose incorporating two variables in the analysis that can improve sensitivity to the toponium signal. Our results indicate that this method may be useful to gain additional insights into the physics phenomenology of the $\ttbar$ threshold region.

\end{abstract}

\begin{keyword}
Toponium Physics

\end{keyword}

\end{frontmatter}




\section{Introduction}

The excess of events observed by the ATLAS and CMS experiments at the Large Hadron Collider (LHC), near the $\ttbar$ pair production threshold in the dileptonic decay mode, is consistent with a spin-0 pseudo-scalar, toponium~\cite{Fadin:1987wz,Fadin:1990wx,CMS:2025dzq,CMS:2025kzt,ATLAS:2026dbe}. CMS first observed~\cite{CMS:2025kzt}, and ATLAS~\cite{ATLAS:2026dbe} confirmed, an excess with a significance exceeding 5 sigma. The reconstruction of the $\ttbar$ dileptonic final state is challenging owing to the presence of two neutrinos in the final state that escape the detector without detection. The ATLAS and CMS experiments employ the Ellipse Method \cite{Betchart:2013nba} and the Sonnenschein Method \cite{Sonnenschein:2005ed,Sonnenschein:2006ud}, respectively for reconstruction in their analyses. We present the use of the Recursive Jigsaw Reconstruction method \cite{Jackson:2017gcy} in this context, as a strategy for distinguishing the toponium signal from the background $\ttbar$ events. 

\section{Monte Carlo Simulation and Event Preselection}
\label{sec1}

The Monte Carlo samples used in this study were generated as per the LHC Run 3 configuration, that is, considering a centre-of-mass (CoM) energy of $\sqrt{s}=13.6 $ TeV. Moreover, all samples were scaled according to the Run 3 integrated luminosity of $\cal{L}$ $= 300 $ fb$^{-1}$. We simulated toponium samples according to the prescription given in ~\cite{Fuks:2024yjj} which reweights matrix elements using Non-Relativistic QCD Green's function. The hard scattered matrix elements are simulated in \textsc{MadGraph5\_aMC@NLO}~\cite{Alwall:2014hca} and \textsc{Pythia8}~\cite{Bierlich:2022pfr} is used for hadronisation. The generation of $\ttbar$ background events follows the use of \textsc{MadGraph5\_aMC@NLO}, \textsc{MadSpin}~\cite{Artoisenet:2012st}, and \textsc{Pythia8}. Two million signal events and four million background events were generated. We set the top quark mass to 173 GeV and the width to $\Gamma_t = 1.49$ GeV.

To reconstruct the jets, we use the anti-$k_T$ algorithm \cite{Cacciari:2008gp} implemented within \textsc{FastJet}~\cite{Cacciari:2011ma} with a radius parameter of $R=0.4$. The \textsc{FastJet} software is accessed via the \textsc{MadAnalysis5}~\cite{Conte:2018vmg,Araz:2020lnp} framework. The normalisations for $\ttbar$ and toponium are $\sigma_{t\Bar{t}} \times Br(W\rightarrow l\nu)^2 = 43.6 ~\rm{pb}$ \cite{ATLAS:2023slx,CMS:2023qyl} and $\sigma_{\eta} \times Br(W\rightarrow l\nu)^2 = 0.277~\rm{pb}$, respectively.

We select events which consist of at least two oppositely charged leptons $(ee,\mu\mu,e\mu)$ and at least two bottom quark initiated jets ($b$-jets). A list of pre-selection criteria applied on the samples are summarised in Table~\ref{tab:preselection}.

\begin{table}[h!]
\centering
\caption{Pre-selection criteria used in this analysis.}
\begin{tabular}{ll}
\hline\hline
\textbf{Selection} & \textbf{Requirement} \\
\hline
b-jet  multiplicity  & $N_{b-jet} \ge 2$ \\
Lepton multiplicity  & $N_{\ell} = 2$, Oppositely charged leptons \\
b-jet $p_T$ & $p_T^{b-jet} > 25$ GeV\\
b-jet $\eta$ & $|\eta|^{b-jet} < 2.5$   \\
Lepton $p_T$ & $p_T^{\ell} > 25$ GeV\\
Lepton $\eta$ & $|\eta|^{\ell} < 2.5$   \\
$\Delta R$($b$-jet,Lepton) & $\Delta R(b\rm{-jet}, \ell) > 0.4$ \\
\hline\hline
\end{tabular}
\label{tab:preselection}
\end{table}

\section{Reconstruction Method for $\ttbar$ in Dilepton Final States}

\begin{figure}[htbp]
    \centering
    \includegraphics[width=0.4\linewidth]{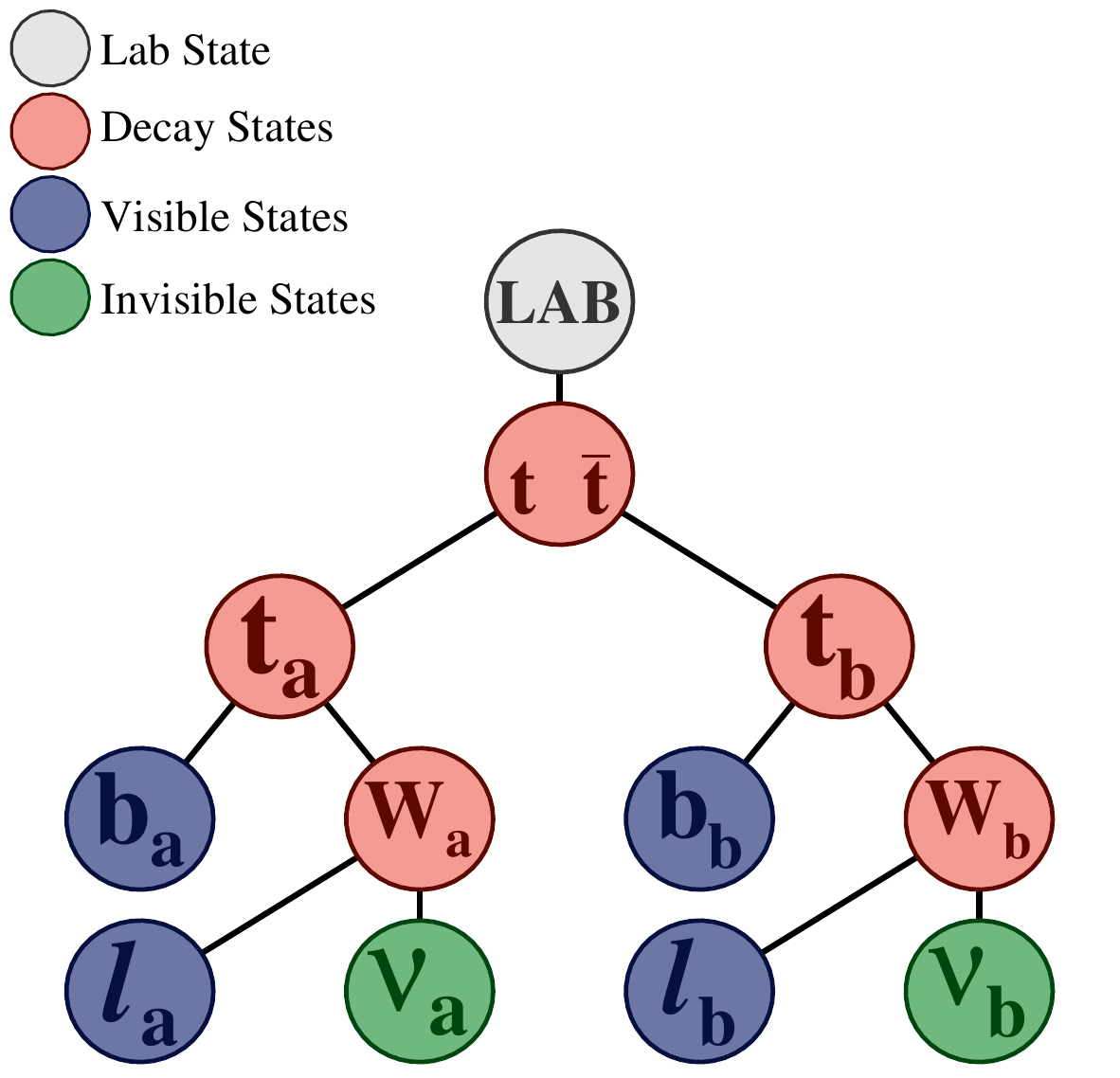}
    \caption{Example decay tree diagram of the process $t\Bar{t} \rightarrow b\Bar{b} W ( l \nu_l) W ( l \Bar{\nu_l})$}
    \label{fig:reco_tree}
\end{figure}

In this study, the reconstruction of $t\bar{t} \rightarrow bbW(l\nu)W(l\nu)$ is implemented in the RestFrames package which is based on the Recursive Jigsaw Reconstruction method~\cite{Jackson:2017gcy}. The reconstruction is challenging owing to both - pairing $b$-jet with the leptons (combinatoric challenge) and to deal with unmeasured particles (two neutrinos). The algorithm first chooses a decay tree that considers all intermediate particle decays with visible and invisible final state particles. Then for each step in the decay tree, it applies a jigsaw rule relating the current frame of reference to the next as a function of kinematic variables. These steps are then applied recursively until it reaches the end of the decay chain. The decay tree used in this analysis is shown in \autoref{fig:reco_tree}. 

\begin{figure}[htbp]
    \centering
    \includegraphics[width=0.6\linewidth]{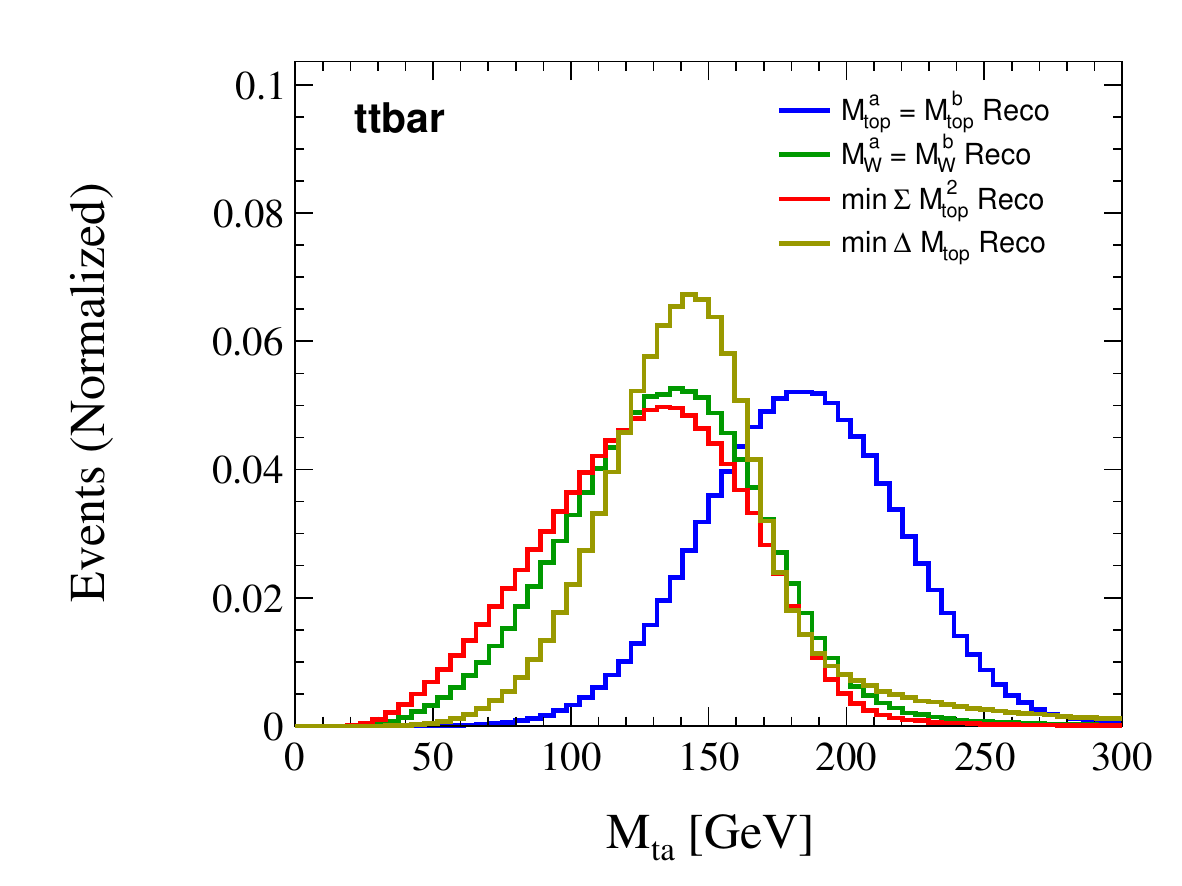}
\caption{Distribution of the $M_{ta}$ variable for the $t\bar{t}$ process.}
\label{fig:mta}
\end{figure}

We apply this Recursive Jigsaw Reconstruction technique \cite{Jackson:2017gcy} via the use of the RestFrames package. There are four different ways in which the dilepton $\ttbar$ system can be constrained. These constraints formulate the following methods:

\begin{enumerate}
    \item[$\textbf{A}$:] $M^a_{\rm top} = M^b_{\rm top}$
    \item[$\textbf{B}$:] $M_{W}^a = M_{W}^b$
    \item[$\textbf{C}$:] min $\Sigma M_{\rm top}^2$
    \item[$\textbf{D}$:] min $\Delta M_{\rm top}$
\end{enumerate}

\begin{figure}[htbp]
    \centering
    \includegraphics[width=0.49\linewidth]{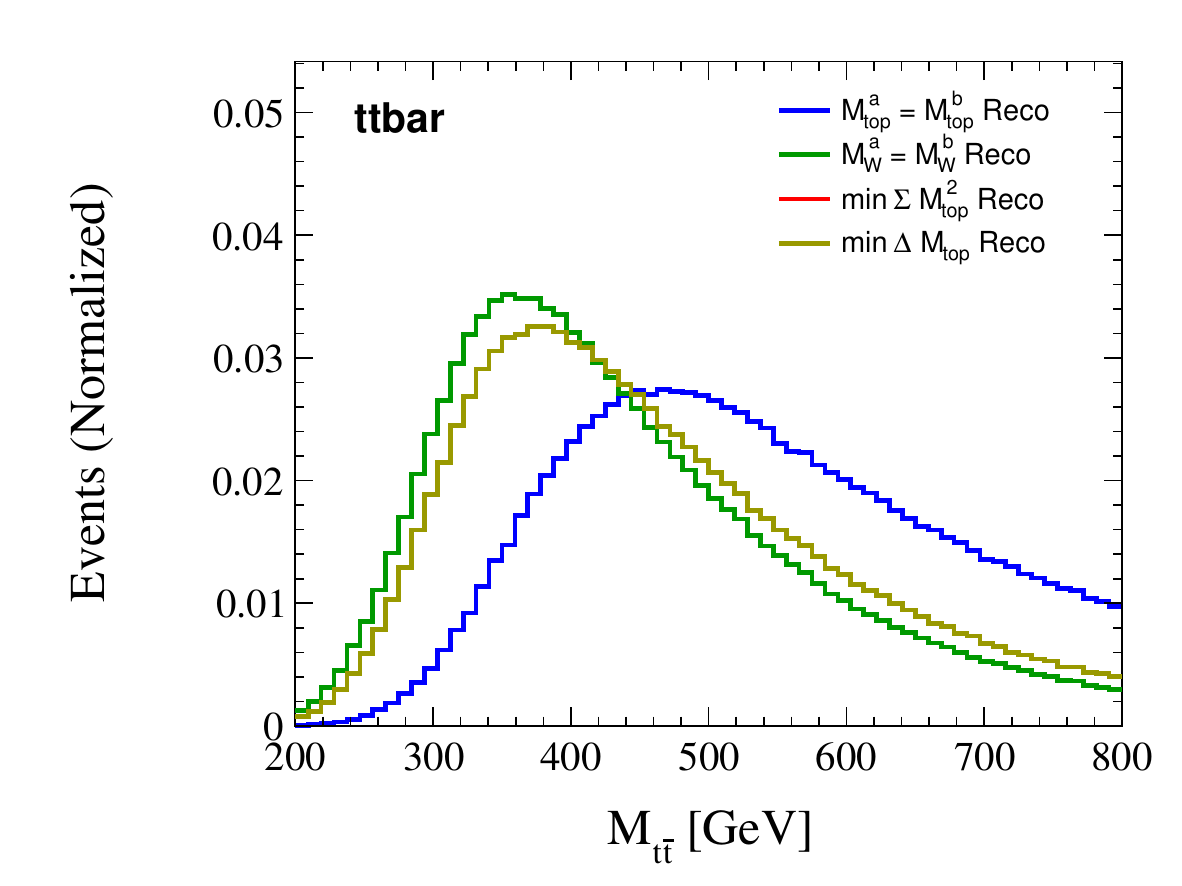}
    \includegraphics[width=0.49\linewidth]{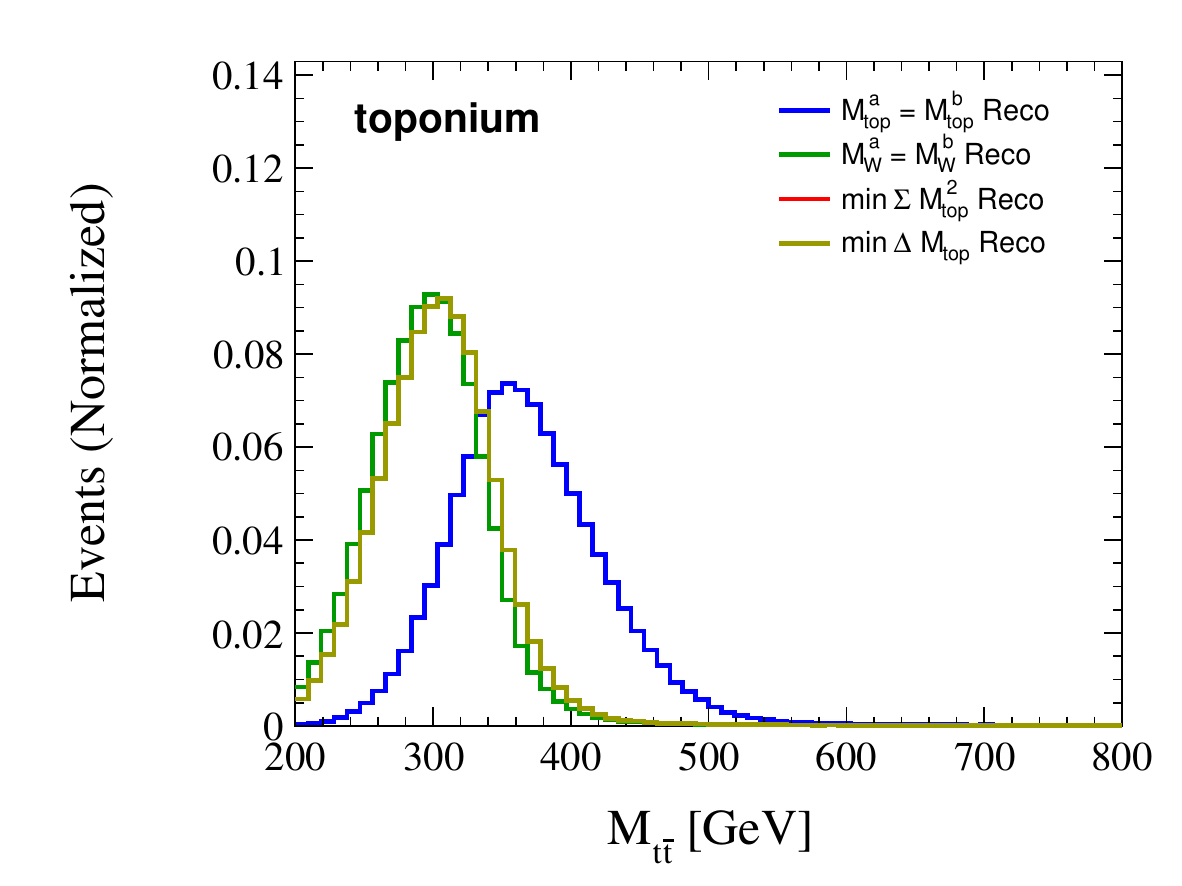}
    \caption{Invariant mass distributions of the top-quark pair as evaluated by the reconstruction algorithms.}
    \label{fig:reco_mtt}
\end{figure}

 As shown in Figure \ref{fig:mta}, reconstruction method A is the optimal method, allowing for a consistent top quark mass. Hence for the remainder of the paper, we use the reconstruction method A. We also present the $M_{\ttbar}$ distribution for the four reconstruction methods in \autoref{fig:reco_mtt}.

\section{Analysis}

\begin{figure}[htbp]
    \centering
    \includegraphics[width=0.49\linewidth]{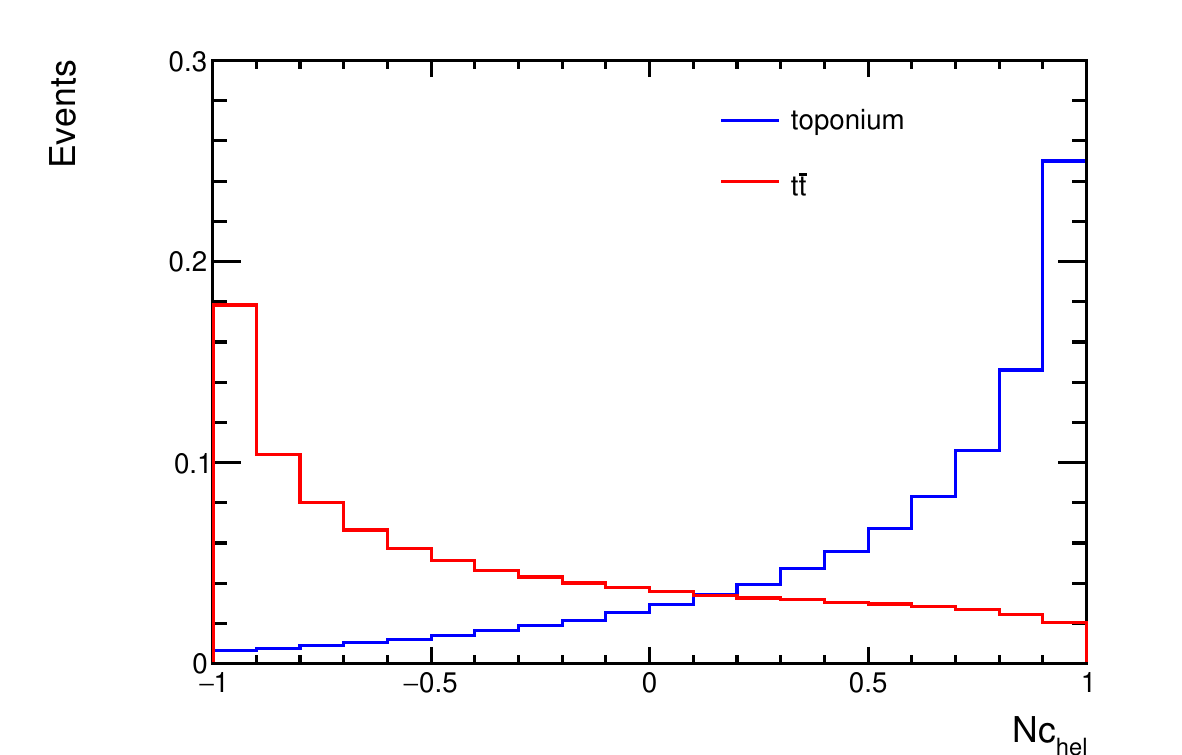}
    \includegraphics[width=0.49\linewidth]{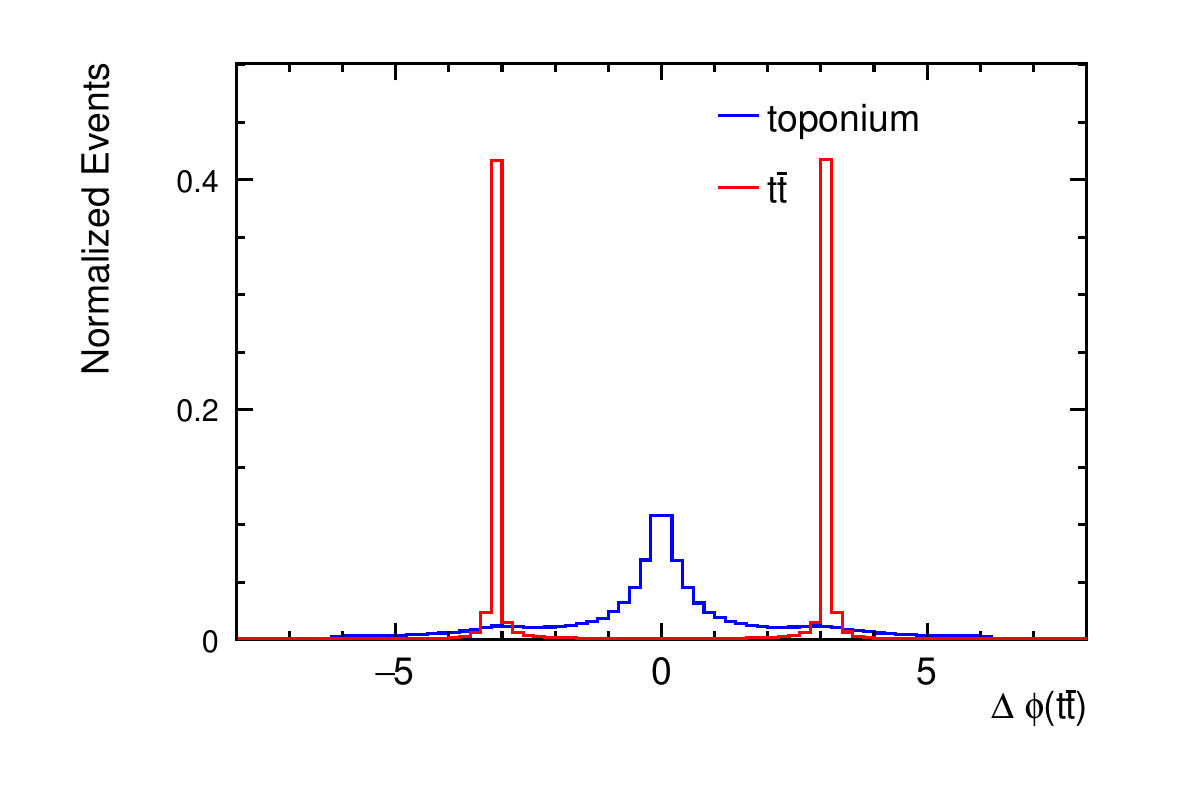}
    \caption{Truth level distributions for $\Delta \phi(t\Bar{t})$ and $N_{chel}$ variables.}
    \label{fig:truth_delta_phi_nchel}
\end{figure}

\begin{figure}[htbp]
    \centering
    \includegraphics[width=0.49\linewidth]{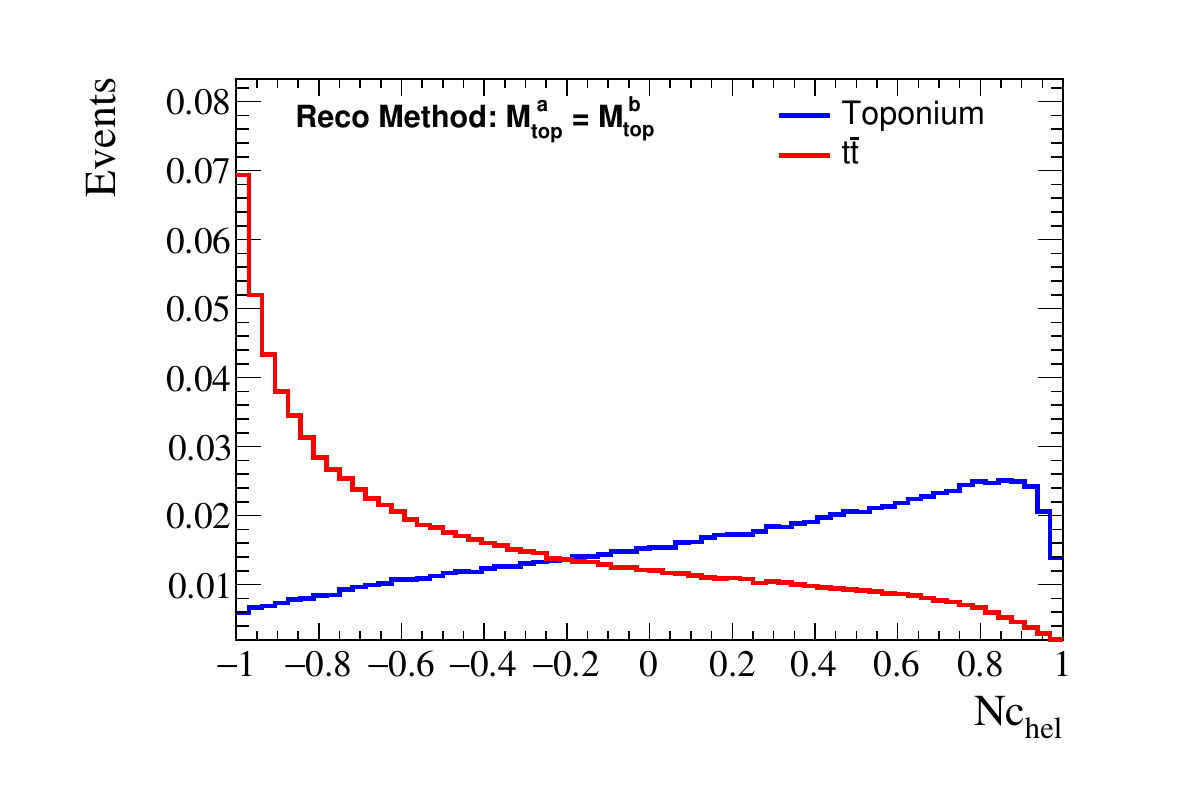}
    \includegraphics[width=0.49\linewidth]{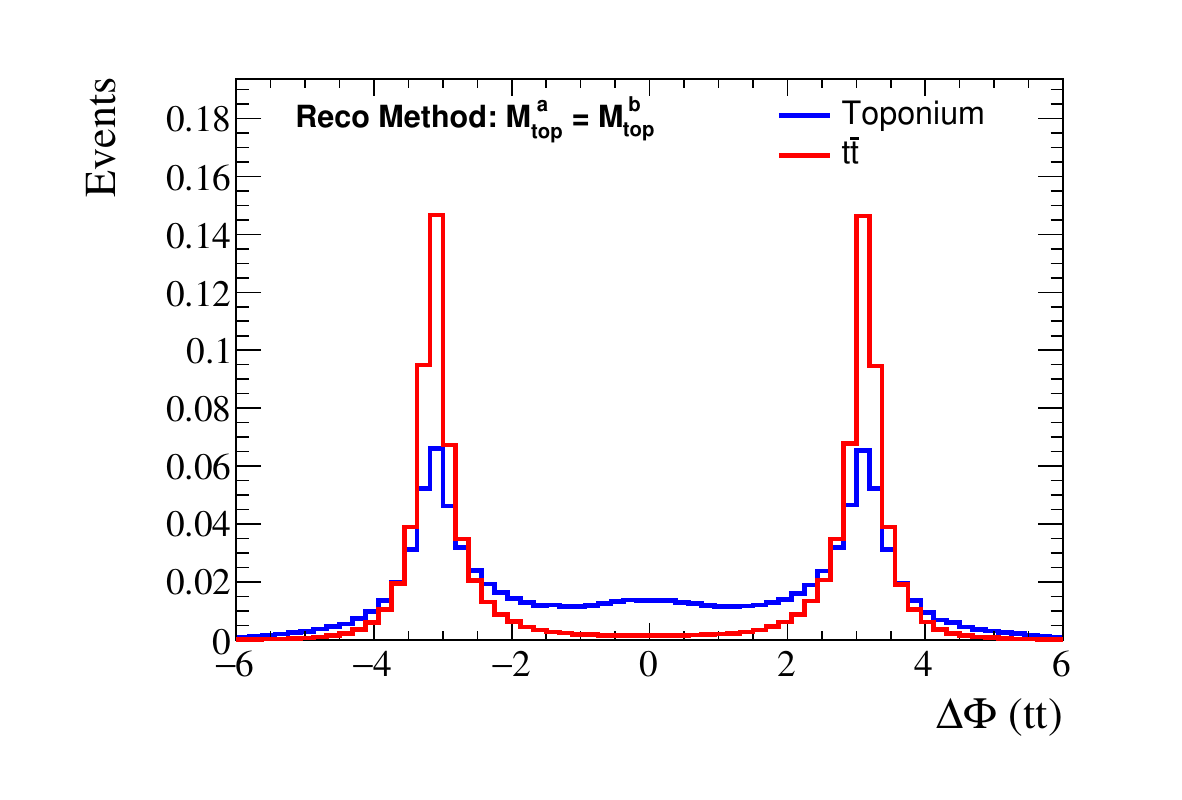}
    \caption{Reconstructed distributions for $\Delta \phi(t\Bar{t})$ and $N_{chel}$ variables obtained by using Method A for reconstruction.}
    \label{fig:reco_delphi_nchell}
\end{figure}

\begin{figure}[htbp]
    \centering
    \includegraphics[width=0.49\linewidth]{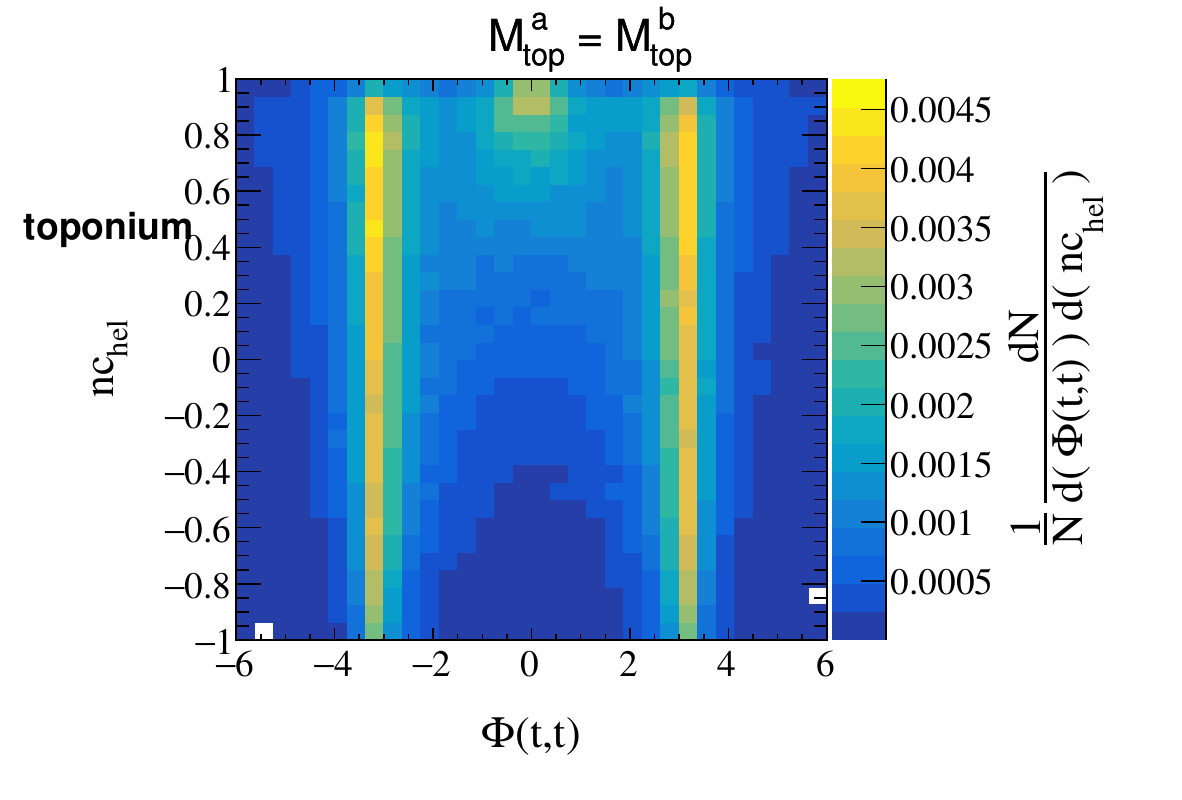}
    \includegraphics[width=0.49\linewidth]{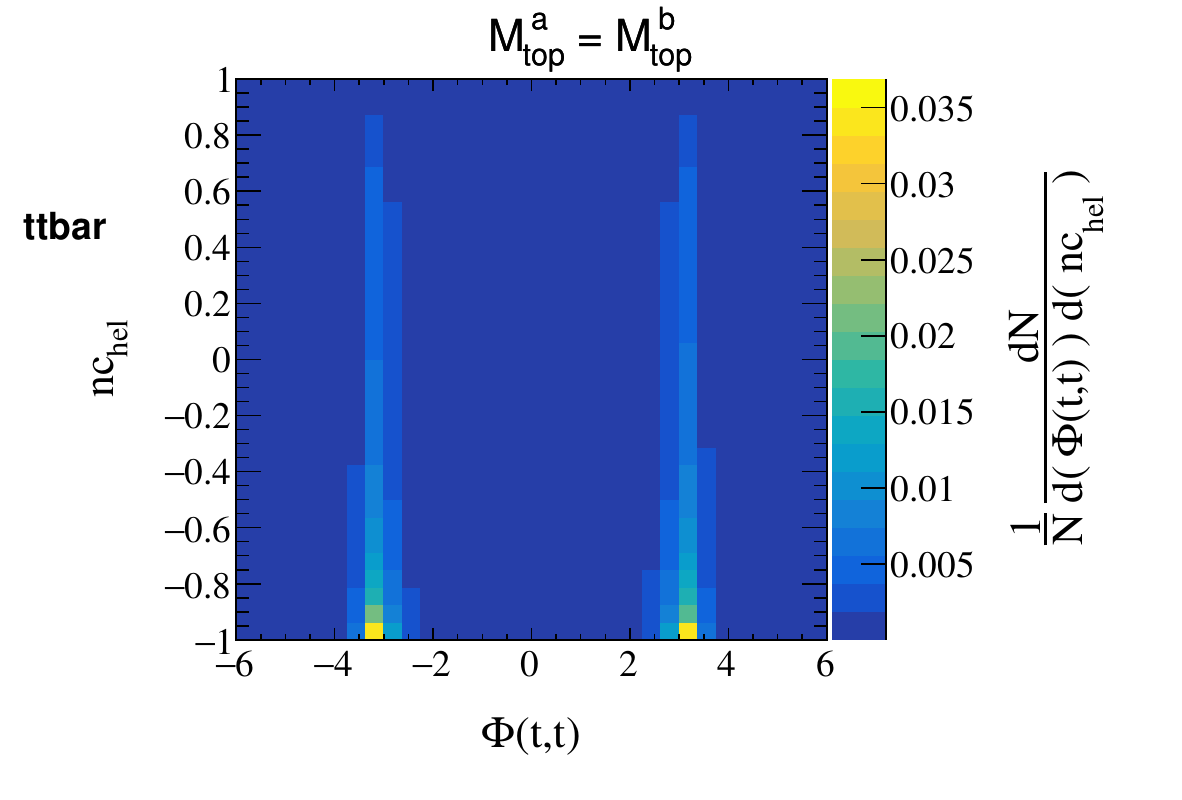}
    \caption{Correlation between the $\Delta \phi(t\Bar{t})$ and $N_{chel}$ variables for Toponium sample (left) and $\ttbar$ sample (right) using method A for reconstruction.}
    \label{fig:corr}
\end{figure}

In this study, we present two variables that can improve the sensitivity for the toponium analysis in the $\ttbar$ dilepton final state. This includes $\Delta \phi(t\Bar{t})$ which is the difference in azimuthal angle of a reconstructed top quark pair, and $\nchel$ which is defined as the scalar product of the lepton's momenta obtained by first boosting the lepton to the $\ttbar$ CoM frame of reference and then boosting it to the parent top-quark frame. The $\nchel$ variable is motivated from the $c_{\rm hel}$ variable~\cite{CMS:2025kzt,ATLAS:2026dbe} but is evaluated in a different frame of reference, in particular in this procedure the top quark is not boosted to the CoM frame.  As shown in Figure \ref{fig:truth_delta_phi_nchel}, the truth-level distributions are presented for these variables for both the $\ttbar$ and toponium samples which are irrespective of the application of reconstruction algorithm. The reconstructed level distributions, obtained after applying reconstruction method A are shown in  \autoref{fig:reco_delphi_nchell}. The correlations between the $\Delta \phi(t\Bar{t})$ and $\nchel$ variables are presented in Figure~\ref{fig:corr} when using method A for reconstruction.  

Nine regions are constructed in the phase space of the $\Delta \phi(t\Bar{t})$ and $N_{chel}$ variables. The bins considered are as follows: 

\begin{align*}
\Delta\Phi(\ttbar) &\in \{[-6,-2],\, [-2,2],\, [2,6]\}; \\
\nchel &\in \{[-1,-0.4],\, [-0.4,0.4],\, [0.4,1]\}; \\[6pt]
\mathcal{R} &= \Delta\Phi(\ttbar) \text{bins} \otimes \nchel\text{bins}.
\end{align*}

\begin{figure}[htbp]
    \centering

    \begin{subfigure}{0.48\linewidth}
        \centering
        \includegraphics[width=.9\linewidth]{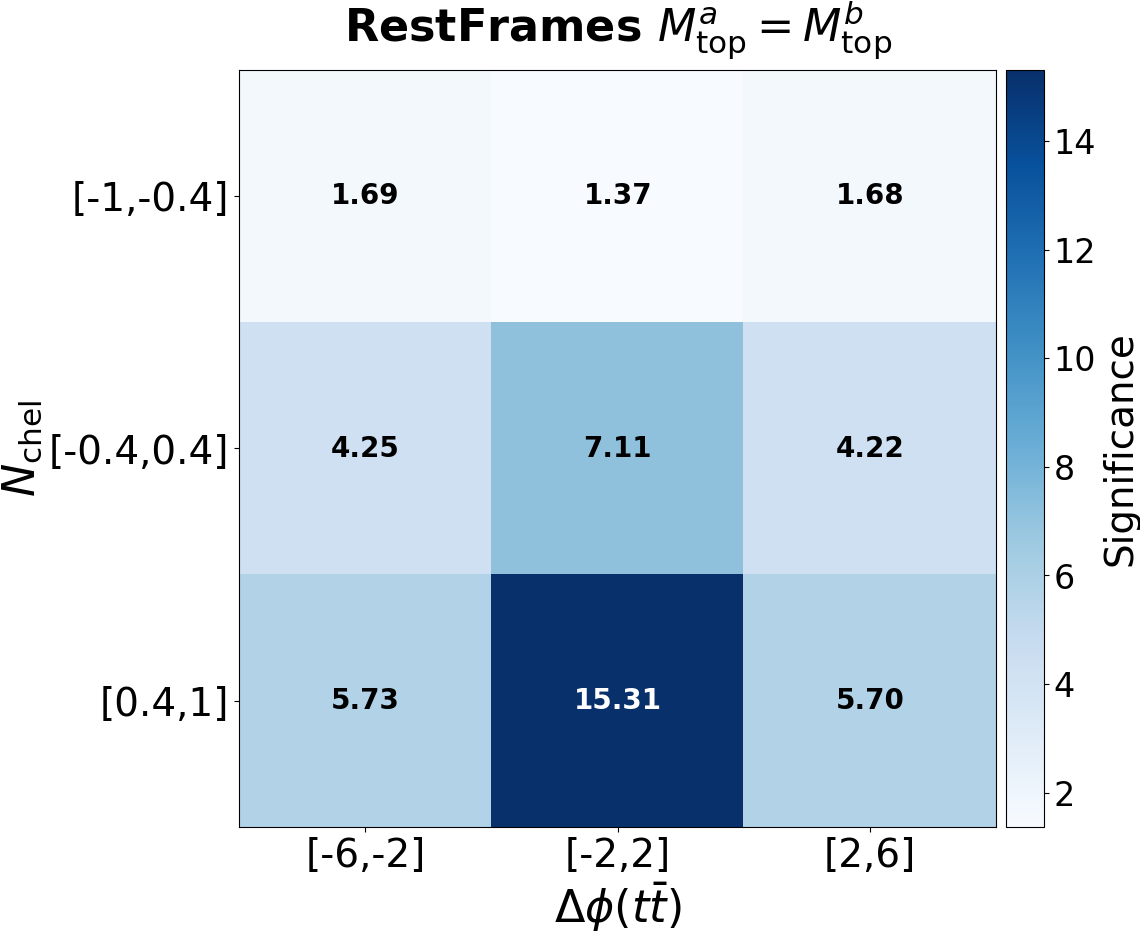}
        \label{fig:nchel_delta_phi_significance}
    \end{subfigure}
    \hfill
    \begin{subfigure}{0.48\linewidth}
        \centering
        \includegraphics[width=.9\linewidth]{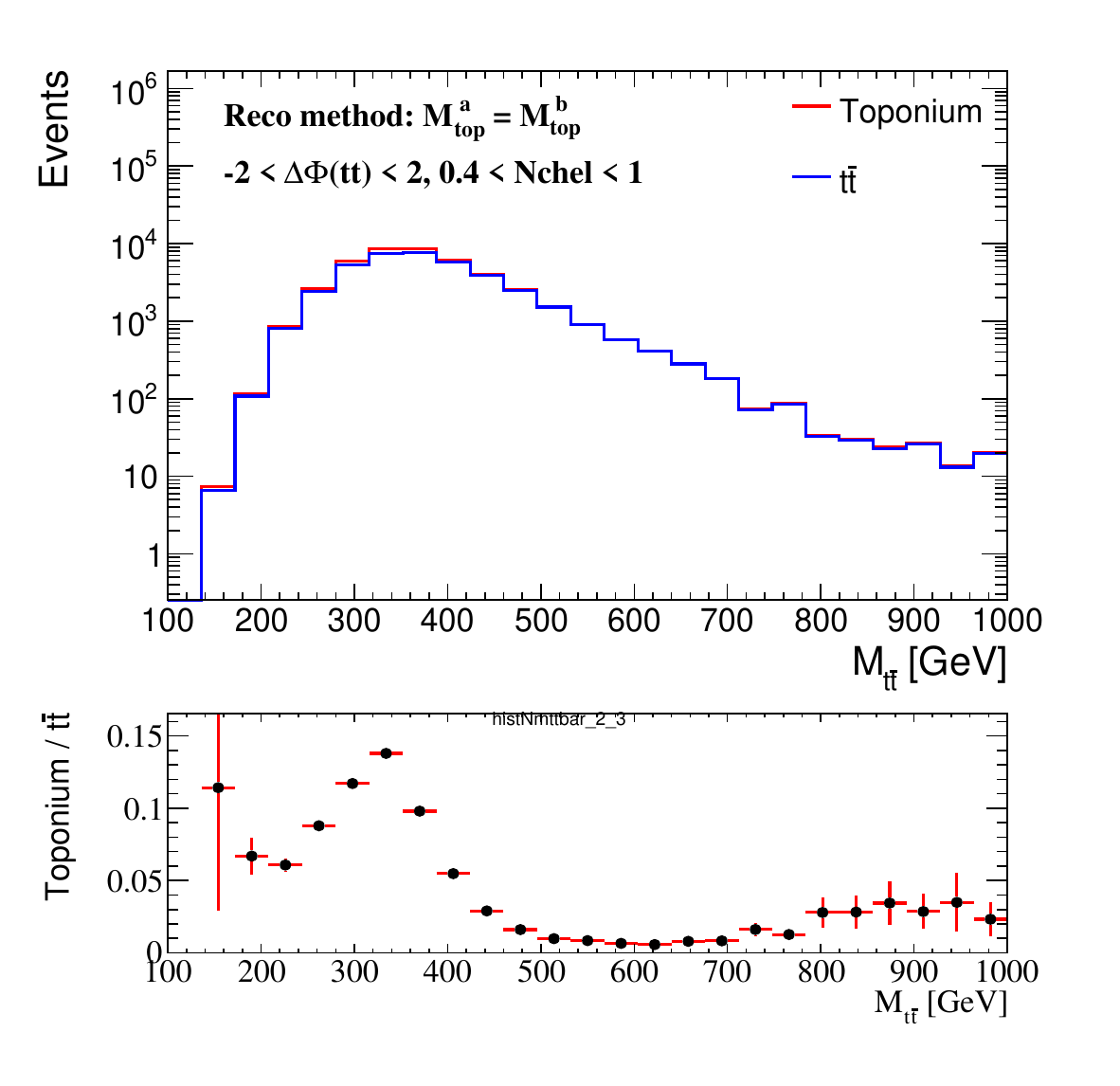}
        \label{fig:reco_mttbar_2_3}
    \end{subfigure}

    \caption{Results after pre-selections using method A for reconstruction. 
    (a) Significance obtained in the nine analysis regions.
    (b) Reconstructed $\ttbar$ invariant mass in the optimal region.}
    \label{fig:significance_and_mttbar}
\end{figure}

The significance metric is defined as $\frac{S}{\sqrt{S+B}}$, where $S$ represents the toponium yield and $B$ represents the $\ttbar$ yield. We use the observable $M_{t\Bar{t}}$ to evaluate the significance, and the results are presented in Figure \ref{fig:significance_and_mttbar}.

As a result of identifying the optimal region defined as $\Delta \phi(t\Bar{t}) \in [-2,2]$ and $N_{chel} \in [0.4,1]$, the $M_{t\Bar{t}}$ distribution is shown in Figure \ref{fig:significance_and_mttbar} and the significance achieved in this region is $15.3\sigma$ without considering systematic uncertainties and detector resolution effects.

\section{Conclusion}

We have presented the Recursive Jigsaw Reconstruction method as an alternative strategy for reconstructing the quasi-bound state toponium, at the Large Hadron Collider, considering the CoM energy of the Run 3 configuration. Within this analysis strategy we have tested two variables, being $N_{chel}$ and $\Delta\phi(t\Bar{t})$, which show that they are effective in discriminating toponium from the $\ttbar$ background. In the phase space defined as $\Delta \phi(t\Bar{t}) \in [-2,2]$ and $N_{chel} \in [0.4,1]$ and without considering systematic uncertainties and detector resolution effects we achieved a significance of $15.3\sigma$ ($\frac{S}{\sqrt{S+B}}$).

\bibliographystyle{unsrt}

\let\oldbibitem\bibitem
\renewcommand{\bibitem}[1]{\oldbibitem{#1}\vspace{-1mm}}

\end{document}